\documentclass[aps,prappl,superscriptaddress,reprint]{revtex4-1}

\usepackage{amsmath}
\usepackage{amssymb}
\usepackage{graphicx}
\usepackage{hyperref}

\begin{document}

\title{Nonreciprocal photon blockade via quadratic optomechanical coupling}
\author{Xun-Wei Xu}
\email{davidxu0816@163.com}
\affiliation{Department of Applied Physics, East China Jiaotong University, Nanchang,
330013, China}
\author{Yan-Jun Zhao}
\affiliation{Beijing National Laboratory for Condensed Matter Physics, Institute of
Physics, Chinese Academy of Sciences, Beijing 100190, China}
\affiliation{School of Physical Sciences, University of Chinese Academy of Sciences,
Beijing 100190, China}
\author{Hui Wang}
\affiliation{Advanced Device Laboratory, RIKEN, Wako, Saitama 351-0198, Japan}
\author{Hui Jing}
\affiliation{Key Laboratory of Low-Dimensional Quantum Structures and Quantum Control of
Ministry of Education, Department of Physics and Synergetic Innovation
Center for Quantum Effects and Applications, Hunan Normal University,
Changsha 410081, China}
\author{Ai-Xi Chen}
\email{aixichen@ecjtu.edu.cn}
\affiliation{Department of Physics, Zhejiang Sci-Tech University, Hangzhou 310018, China}
\affiliation{Department of Applied Physics, East China Jiaotong University, Nanchang,
330013, China}
\date{\today }

\begin{abstract}
We propose to manipulate the statistic properties of the photons transport
nonreciprocally via quadratic optomechanical coupling. We present a scheme
to generate quadratic optomechanical interactions in the normal optical
modes of a whispering-gallery-mode (WGM) optomechanical system by
eliminating the linear optomechanical couplings via anticrossing of
different modes. By optically pumping the WGM optomechanical system in one
direction, the effective quadratic optomechanical coupling in that direction
will be enhanced significantly, and nonreciprocal photon blockade will be
observed consequently. Our proposal has potential applications for the on-chip
nonreciprocal single-photon devices.
\end{abstract}

\maketitle



\section{Introduction}

Nonreciprocal devices~\cite{JalasNPT13}, such as isolators and circulators,
have drawn an immense amount of interest in the past few years, for their
irreplaceable role in signal processing and communication. One of the key
parameters for nonreciprocal devices is the isolation, and almost all the
studies on nonreciprocity focus on the transmission properties of the
nonreciprocal devices. However, whether the statistic properties of the
transmitted photons have been changed and how to manipulate the statistic
properties of the transmitted photons in nonreciprocal devices are rarely
discussed.

Recently, the statistic properties of the transmitted photons in rotating
nonlinear devices were discussed theoretically~\cite{RHuangarX18}, and a
quantum effect called nonreciprocal photon blockade was predicted, that
photon blockade can emerge when the resonator is driven in one direction but
not the other. Physically, the nonreciprocal photon blockade is induced by
the Fizeau-Sagnac drag~\cite{MalykinPU00,HLv17,MaayaniNat18}, which leads to
a split of the resonance frequencies of the counter-circulating modes.
Similarly, nonreciprocal transport~\cite{DWWangPRL13,HorsleyPRL13} and
localization~\cite{RamezaniPRL18} of photons have been demonstrated based on
the Doppler shift in moving photonic lattice made. In contrast to the
classical nonreciprocal behaviors, purely quantum effects in nonreciprocal
devices were proposed in reference~\cite{RHuangarX18}, which opens up the
prospect of exploring nonreciprocal quantum effects, such as nonreciprocal
single-photon blockades, nonreciprocal two-photon blockades, and
nonreciprocal photon-induced tunneling.

An optomechanical system, that resonance frequency of a cavity mode depends
on the position a mechanical mode via radiation pressure or optical gradient
forces, provides us an appropriate platform to manipulate photons (for
reviews, see Refs.~\cite%
{KippenbergSci08,MarquardtPhy09,AspelmeyerPT12,AspelmeyerARX13,MetcalfeAPR14,YLLiuCPB18}%
). Lately, several theoretical~\cite%
{ManipatruniPRL09,HafeziOE12,SchmidtOpt15,MetelmannPRX15,XuXWPRA15,FangKArx15,XWXuPRA16a,MetelmannarX16a,LTianarx16a,MiriarX16a,GLiPRA18}
and experimental~\cite%
{KimNPy15,CHDongNC15,ZShenNPn16,RuesinkNC16a,KFangNPy17a,BernierarX16a,PetersonarX17,BarzanjeharX17,HQiuOE17,ZShenNC18,RuesinkNC18}
works have demonstrated that optomechanical interaction can lead to
nonreciprocal transport of photons. One of the proposals for nonreciprocity
is based on the inherent non-trivial topology in whispering-gallery-mode
(WGM) optomechanical system~\cite{HafeziOE12}, where the effective
optomechanical coupling is enhanced in one direction and suppressed in the
other one by optically pumping the ring resonator. But the enhanced
effective optomechanical coupling is a simple bilinear interaction, which
can not be used to manipulate the statistic properties of the nonreciprocal
transport photons. Fortunately, Xie \emph{et al.} proposed to generate
strong quadratic (nonlinear) optomechanical coupling by a strong driving
optical field, and the appearance of strong photon antibunching was
predicted in a quadratically coupled optomechanical system under
single-photon weak coupling conditions~\cite{HXiePRA17}.

Motivated by the pioneering work on nonreciprocal quantum effects~\cite%
{RHuangarX18}, we propose to manipulate the statistic properties of the
nonreciprocal transport photons in a WGM optomechanical system~\cite%
{HafeziOE12} with quadratic optomechanical interactions~\cite{HXiePRA17}.
Employing a similar idea given in Refs.~\cite%
{HeinrichPRA10,HZWuNJP13,JTHill13,ParaisoPRX15}, quadratic optomechanical
interactions can be generated in the normal optical modes of a WGM
optomechanical system by eliminating the linear optomechanical couplings via
anticrossing of different modes. We demonstrate that quadratic
optomechanical interactions can not only induce nonreciprocal photon
transport, but also manipulate the statistic properties of the nonreciprocal
transport photons. For example, photon blockade with high transmission
coefficient can be observed when the photons transport in one direction but
not the other. WGM optomechanical systems with quadratic optomechanical
interactions can be used to design nonreciprocal single-photon devices in
integrated photonic chips.

The remainder of this paper is organized as follows. In Sec.~II, we propose
an system consisting of two whispering gallery mode (WGM) resonators quadratically coupling with
a common mechanical mode. In Sec.~III, the effective Hamiltonian of
an optomechanical system with quadratic coupling is obtained with one
optical mode driven by a strong external field. In Sec.~IV, we show that the
optomechanical system with quadratic coupling can be used to realize
nonreciprocal photon blockade. Finally, the main results are summarized in
Sec.~V.

\section{Quadratic optomechanical coupling}

\begin{figure}[tbp]
\includegraphics[bb=78 175 545 637, width=8.5 cm, clip]{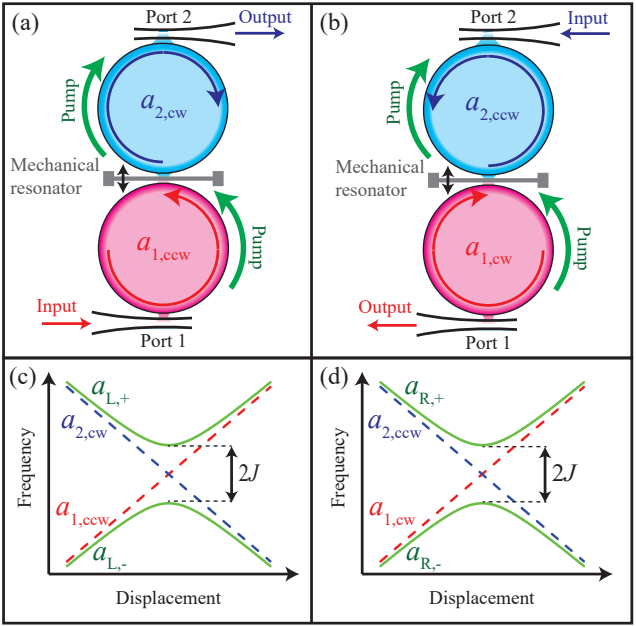}
\caption{(Color online) (a) and (b) Schematic diagram for generating
quadratic optomechanical coupling, where a mechanical nanostring oscillator
is placed between two whispering gallery mode (WGM) resonators. (c) and (d)
Dispersion of the optical modes as a function of the displacement.}
\label{fig1}
\end{figure}

One of the most major ingredients in optomechanical system is that the
resonance frequency of a cavity mode is dependent on the position of a
mechanical mode. It is well known that the cavity frequency of a WGM
optomechanical system is almost linearly proportional to the the mechanical
position~\cite{AnetsbergerNPy09}, even though the effects of quadratic
optomechanical coupling have also been observed in experiments~\cite%
{DoolinPRA14,BrawleyNC16}. In this section we will show how to generate a
quadratic optomechanical coupling as the dominant coupling in the situation
when the linear optomechanical coupling vanishes.

As shown in Fig.~\ref{fig1}(a) and \ref{fig1}(b), the setup we consider here consists of one mechanical resonator optomechanical
coupling to two optical resonators ($j=1,2$) via the optical evanescent
field, with each optical
resonator supporting two degenerate clockwise (CW) and counter-clockwise (CCW)
travelling-wave whispering-galley modes (WGMs). This model can be described by
the optomechanical interaction Hamiltonian
\begin{eqnarray}
H_{\mathrm{om}} &=&\sum_{j=1,2}\sum_{\lambda =\mathrm{cw,ccw}}\left[ \omega
_{0}+\left( -1\right) ^{j}g_{0}q\right] a_{j,\lambda }^{\dag }a_{j,\lambda }
\notag \\
&&+J\left( a_{1,\mathrm{ccw}}a_{2,\mathrm{cw}}^{\dag }+a_{1,\mathrm{cw}}a_{2,%
\mathrm{ccw}}^{\dag }+\mathrm{H.c.}\right)  \notag \\
&&+\frac{1}{2}\omega _{m}\left( q^{2}+p^{2}\right) ,
\end{eqnarray}%
where $a_{j,\lambda }$ and $a_{j,\lambda }^{\dag }$ ($j=1,2$ and $\lambda =%
\mathrm{cw,ccw}$) are the annihilation and creation operators of the optical
modes with frequency $\omega _{0}$; $J$ is the tunneling amplitude between
the optical modes; $q$ and $p$ are the dimensionless displacement and
momentum operators of the mechanical resonator with frequency $\omega _{m}$,
and $g_{0}$ is the linear optomechanical coupling strength between the
mechanical resonator and optical modes.
We assume that the optical mode $a_{j,\lambda }$ is coupled to a waveguide (Port $j$) with strength $\gamma _{c}$, and and the damping rate of mechanical resonator $q$ is $\gamma _{m}$.

Following the approach in Refs.~\cite%
{HeinrichPRA10,HZWuNJP13,JTHill13,ParaisoPRX15}, where $\left\vert J\right\vert
\gg \omega _{m}$ is assumed such that $q$ can be treated as a quasi-static
variable, the Hamiltonian can be diagonalized as
\begin{equation}
H_{\mathrm{om}}=\sum_{j=L,R}\sum_{\lambda =\pm }\omega _{\lambda }\left(
q\right) a_{j,\lambda }^{\dag }a_{j,\lambda }+\frac{1}{2}\omega _{m}\left(
q^{2}+p^{2}\right) ,
\end{equation}%
in the normal modes basis, $a_{L,\pm }=[ Ja_{1,\mathrm{ccw}}+(
g_{0}q\pm \sqrt{J^{2}+( g_{0}q) ^{2}}) a_{2,\mathrm{cw}}%
] /D_{\pm }$ and $a_{R,\pm }=[ Ja_{1,\mathrm{cw}}+(
g_{0}q\pm \sqrt{J^{2}+( g_{0}q) ^{2}}) a_{2,\mathrm{ccw}}%
] /D_{\pm }$, with $D_{\pm }^{2}=J^{2}+( g_{0}q\pm \sqrt{%
J^{2}+( g_{0}q) ^{2}}) ^{2}$, and eigenfrequencies%
\begin{equation}
\omega _{\pm }\left( q\right) =\omega _{0}\pm \sqrt{J^{2}+\left(
g_{0}q\right) ^{2}}
\end{equation}%
as shown in Figs.~\ref{fig1}(c) and \ref{fig1}(d). Moreover, $\left\vert
J\right\vert \gg g_{0}q$ is assumed such that we can Taylor expand the
eigenfrequencies as
\begin{equation}
\omega _{\pm }\left( q\right) \approx \omega _{\pm }\pm \frac{g_{0}^{2}}{2J}%
q^{2}
\end{equation}%
with frequencies $\omega _{\pm }\equiv \omega _{\pm }\left( 0\right) =\omega
_{0}\pm J$, and the quasi-static Hamiltonian, with quadratic optomechanical
coupling $g\equiv g_{0}^{2}/(2J)$ between the mechanical resonator and
quasi-static normal optical modes, $a_{L,\pm }\approx\left( a_{1,\mathrm{ccw}}\pm
a_{2,\mathrm{cw}}\right) /\sqrt{2}$ and $a_{R,\pm }\approx\left( a_{1,\mathrm{cw}%
}\pm a_{2,\mathrm{ccw}}\right) /\sqrt{2}$, is given approximately by
\begin{eqnarray}
H_{\mathrm{om}} &\approx &\left( \omega _{+}+gq^{2}\right) \left(
a_{L,+}^{\dag }a_{L,+}+a_{R,+}^{\dag }a_{R,+}\right)   \notag \\
&&+\left( \omega _{-}-gq^{2}\right) \left( a_{L,-}^{\dag
}a_{L,-}+a_{R,-}^{\dag }a_{R,-}\right)   \notag \\
&&+\frac{1}{2}\omega _{m}\left( q^{2}+p^{2}\right) .
\end{eqnarray}

As already shown in the experiment~\cite{ParaisoPRX15}, when the the tunneling amplitude between
the optical modes $J$ is larger than the optical damping rates $\gamma _{c}$, the transmission spectrum of a laser probe through the coupled optical modes features resonance dips at the normal resonance frequencies $\omega _{\pm }$, not at the bare optical resonance frequencies $\omega _{0}$. That is to say, the normal modes are coupled to the external waveguides and can be used to describe the input-output characteristic of the coupled optical modes system. Specifically, the total loss damping rate of the normal modes $a_{L/R,\pm}$ is $\gamma _{c}$ for $a_{R,\pm }\approx\left[ a_{1,\mathrm{cw}%
}\pm a_{2,\mathrm{ccw}}\right) /\sqrt{2}$, and they are coupled to both the two ports with strength $\gamma _{c}/2$, respectively.

\section{Directional nonlinear interaction}

In this section, we choose either pair of degenerated quasi-static normal
optical modes ($a_{L,+}$ and $a_{R,+}$, or $a_{L,-}$ and $a_{R,-}$), to
generate strong nonlinear interaction for few photons traveling in one
direction, but not in the reverse direction. Without loss of generality, the
two optical modes are denoted as $a_{L}$ and $a_{R}$, with frequency $\omega
_{a}=\omega _{+}$ or $\omega _{-}$, for photons travelling from port $1$ to
port $2$ and the opposite direction, respectively. 
To enhance the nonlinear optomechanical coupling between the optical mode $%
a_{L}$ and the mechanical resonator, the optical mode $a_{L}$ is pumped by a
strong field with amplitude $\Omega \gg \gamma _{c}$ and frequency $\omega
_{L} \sim \omega _{a} - 2\omega _{m}$. In the rotating reference frame with
the optical frequency $\omega _{L}$, the system can be described by a
Hamiltonian as
\begin{eqnarray}
H &=&\left( \Delta _{a}+gq^{2}\right) \left( a_{L}^{\dag }a_{L}+a_{R}^{\dag
}a_{R}\right)  \notag \\
&&+\frac{1}{2}\omega _{m}\left( q^{2}+p^{2}\right)  \notag \\
&&+\Omega a_{L}^{\dag }+\Omega a_{L},
\end{eqnarray}%
with detuning $\Delta _{a}=\omega _{a}-\omega _{d}$.

Under strong driving condition, we perform the displacement transformations: $%
a_{L}\rightarrow \alpha _{L}+a_{L}$, $a_{R}\rightarrow \alpha_{R}+a_{R}$, $%
q\rightarrow q_{s}+q$ and $p\rightarrow p_{s}+p$, where $\alpha _{L}$, $%
\alpha _{R}$, $q_{s}$ and $p_{s}$ are the steady state values, and $a_{L}$, $%
a_{R}$, $q$, and $p$ (on the right side of the arrow) are the quantum
fluctuation operators. The steady state values $\alpha _{L}$, $\alpha _{R}$,
$q_{s}$ and $p_{s}$ can be obtained by the equations of motions yielding $\alpha
_{L}=-i2\Omega /(\gamma _{c}+i2\Delta _{a})$, and $\alpha _{R}=q_{s}=p_{s}=0$%
. The operators $q$ and $p$ for the mechanical resonator can be written in
terms of phonon creation and annihilation operators as $q=\left( b^{\dag
}+b\right) /\sqrt{2}$, $p=i\left( b^{\dag }-b\right) /\sqrt{2}$, and the
effective Hamiltonian for the quantum fluctuation operators reads
\begin{eqnarray}
H_{\mathrm{eff}} &=&\Delta _{a}a_{L}^{\dag }a_{L}+\Delta _{a}a_{R}^{\dag
}a_{R}+\omega _{m}b^{\dag }b  \notag \\
&&+\frac{g}{2}\left( \left\vert \alpha \right\vert ^{2}+a_{L}^{\dag
}a_{L}+a_{R}^{\dag }a_{R}\right) \left( b^{\dag }+b\right) ^{2}  \notag \\
&&+\frac{g}{2}\left( \alpha_L a^{\dag }+\alpha ^{\ast }a\right) \left(
b^{\dag }+b\right) ^{2}.
\end{eqnarray}%
We assume that the optical driving field is strong enough so the steady
state value $\alpha_L$ is much larger than the quantum fluctuation operators
$a$, i.e., $\vert \alpha_L \vert ^{2}\gg \langle a_L^{\dag
}a_L\rangle \sim \langle a_R^{\dag }a_R\rangle $, and the
term $g( a_{L}^{\dag }a_{L}+a_{R}^{\dag }a_{R}) ( b^{\dag
}+b) ^{2}$ in the above equation can be neglected safely. For $\Delta
_{a} \sim 2 \omega _{m} \gg |g\alpha _{L}|/2$, the effective Hamiltonian can
be further simplified by rotating-wave approximation and neglecting
the high frequency terms $b^2$ and $ab^2$ yielding
\begin{eqnarray}
H_{\mathrm{eff}} &=&\Delta _{a}\left( a_{L}^{\dag }a_{L}+a_{R}^{\dag
}a_{R}\right) +\omega _{m}^{\prime }b^{\dag }b  \notag \\
&&+Ga_{L}^{\dag }b^{2}+G^{\ast }a_{L}b^{\dag 2},
\end{eqnarray}
where $\omega _{m}^{\prime }=\omega _{m}+g\left\vert \alpha\right\vert ^{2}$
is the effective mechanical frequency, and $G=g\alpha _{L}/2$ is the effective
nonlinear coupling strength between the optical and mechanical modes.
Without loss of generality $G$ is assumed to be real in the following.

To investigate the system's response behavior to weak prob fields, a weak
field with amplitude $\varepsilon \ll \gamma _{c}$ and frequency $\omega
_{p} \approx \omega _{a}$ is input from one of the ports. The total
Hamiltonian is given by
\begin{equation}
H_{\mathrm{tot}}=H_{\mathrm{eff}}+H_{\mathrm{probe}},
\end{equation}
where $H_{\mathrm{probe}}$ describes the probe field. If the probe field is
input from port $1$, it can be described by
\begin{equation}
H_{\mathrm{probe}}=\varepsilon e^{-i\delta t}a_{L}^{\dag }+\mathrm{H.c.},
\end{equation}%
while if the weak field is input from port $2$, it can be given by
\begin{equation}
H_{\mathrm{probe}}=\varepsilon e^{-i\delta t}a_{R}^{\dag }+\mathrm{H.c.},
\end{equation}%
where $\delta=\omega_p-\omega_d$ is the detuning between the strong driving
and weak probe fields. In the rotating reference frame with the unitary
operator $R^{\prime }\left( t\right) =\mathrm{exp}[ i\delta (
a_{L}^{\dag }a_{L}+a_{R}^{\dag }a_{R}+b^{\dag }b/2) t] $, $H_{%
\mathrm{probe}}$ becomes time independent, and the effective Hamiltonian
becomes%
\begin{eqnarray}
H_{\mathrm{eff}} &=&\Delta a_{L}^{\dag }a_{L}+\Delta a_{R}^{\dag
}a_{R}+\Delta _{m}b^{\dag }b  \notag \\
&&+Ga_{L}^{\dag }b^{2}+G^{\ast }a_{L}b^{\dag 2},
\end{eqnarray}
where the detunings $\Delta =\Delta _{a}-\delta $ and $\Delta _{m}=\omega
_{m}+g\left\vert \alpha \right\vert ^{2}-\delta /2$ satisfy the condition $%
\max \left\{ \left\vert \Delta \right\vert ,\left\vert \Delta _{m}\right\vert
\right\} \ll \omega _{m}$.

According to the input-output relations~\cite{GardinerPRA85}, we have $a_{1,%
\mathrm{in}}=\varepsilon /\sqrt{\gamma _{c}/2}$ and $a_{2,\mathrm{out}}=%
\sqrt{\gamma _{c}/2}a_{L}$ ($a_{2,\mathrm{in}}=\varepsilon /\sqrt{\gamma
_{c}/2}$ and $a_{1,\mathrm{out}}=\sqrt{\gamma _{c}/2}a_{R}$), and then the
transmission coefficient for the weak probe field can be defined by
\begin{equation}
T_{21}\equiv \frac{\left\langle a_{2,\mathrm{out}}^{\dag }a_{2,\mathrm{out}%
}\right\rangle }{\left\langle a_{1,\mathrm{out}}^{\dag }a_{1,\mathrm{out}%
}\right\rangle }=\frac{\gamma _{c}^{2}}{4\varepsilon }\left\langle
a_{L}^{\dag }a_{L}\right\rangle
\end{equation}%
for photon transport from port $1$ to port $2$, and
\begin{equation}
T_{12}\equiv \frac{\left\langle a_{1,\mathrm{out}}^{\dag }a_{1,\mathrm{out}%
}\right\rangle }{\left\langle a_{2,\mathrm{out}}^{\dag }a_{2,\mathrm{out}%
}\right\rangle }=\frac{\gamma _{c}^{2}}{4\varepsilon }\left\langle
a_{R}^{\dag }a_{R}\right\rangle
\end{equation}%
for photon transport from port $2$ to port $1$, where $n_{L}\equiv
\left\langle a_{L}^{\dag }a_{L}\right\rangle $ and $n_{R}\equiv \left\langle
a_{R}^{\dag }a_{R}^{\dag }\right\rangle $ are the mean photon numbers. The
isolation for probe field transport from port $1$ to port $2$ is defined by
\begin{equation}
I \equiv \frac{T_{12}}{T_{12}}.
\end{equation}

Using the input-output relations: $a_{1,\mathrm{in}}=\varepsilon /\sqrt{%
\gamma _{c}/2}$ and $a_{2,\mathrm{out}}=\sqrt{\gamma _{c}/2}a_{L}$ ($a_{2,%
\mathrm{in}}=\varepsilon /\sqrt{\gamma _{c}/2}$ and $a_{1,\mathrm{out}}=%
\sqrt{\gamma _{c}/2}a_{R}$), the statistic properties of the transmitted
photons $a_{2,\mathrm{out}}$ and $a_{1,\mathrm{out}}$ can be described by
the second-order correlation functions in the steady state ($t\rightarrow
\infty $)
\begin{eqnarray}
g_{21}^{\left( 2\right) }\left( \tau \right) &\equiv &\frac{\left\langle
a_{2,\mathrm{out}}^{\dag }\left( t\right) a_{2,\mathrm{out}}^{\dag }\left(
t+\tau \right) a_{2,\mathrm{out}}\left( t+\tau \right) a_{2,\mathrm{out}%
}\left( t\right) \right\rangle }{\left\langle a_{2,\mathrm{out}}^{\dag
}\left( t\right) a_{2,\mathrm{out}}\left( t\right) \right\rangle ^{2}}
\notag \\
&=&\frac{\left\langle a_{L}^{\dag }\left( t\right) a_{L}^{\dag }\left(
t+\tau \right) a_{L}\left( t+\tau \right) a_{L}\left( t\right) \right\rangle
}{\left\langle a_{L}^{\dag }\left( t\right) a_{L}\left( t\right)
\right\rangle ^{2}}
\end{eqnarray}%
for photon transport from port $1$ to port $2$, and
\begin{eqnarray}
g_{12}^{\left( 2\right) }\left( \tau \right) &\equiv &\frac{\left\langle
a_{1,\mathrm{out}}^{\dag }\left( t\right) a_{1,\mathrm{out}}^{\dag }\left(
t+\tau \right) a_{1,\mathrm{out}}\left( t+\tau \right) a_{1,\mathrm{out}%
}\left( t\right) \right\rangle }{\left\langle a_{1,\mathrm{out}}^{\dag
}\left( t\right) a_{1,\mathrm{out}}\left( t\right) \right\rangle ^{2}}
\notag \\
&=&\frac{\left\langle a_{R}^{\dag }\left( t\right) a_{R}^{\dag }\left(
t+\tau \right) a_{R}\left( t+\tau \right) a_{R}\left( t\right) \right\rangle
}{\left\langle a_{R}^{\dag }\left( t\right) a_{R}\left( t\right)
\right\rangle ^{2}}
\end{eqnarray}%
for photon transport from port $2$ to port $1$.

In the next section, the transmission coefficients and correlation functions
will be obtained by numerically solving the master equation for the density
matrix $\rho $ of the system~\cite{Carmichael93}%
\begin{eqnarray}
\frac{\partial \rho }{\partial t} &=&-i\left[ H_{\mathrm{tot}},\rho \right]
+\gamma _{c}L[a_{L}]\rho +\gamma _{c}L[a_{R}]\rho  \notag  \label{Eq6} \\
&&+\gamma _{m}\left( n_{\mathrm{th}}+1\right) L[b]\rho +\gamma _{m}n_{%
\mathrm{th}}L[b^{\dag }]\rho,
\end{eqnarray}%
where $L[o]\rho =o\rho o^{\dag }-\left( o^{\dag }o\rho +\rho o^{\dag
}o\right) /2$ denotes a Lindbland term for an operator $o$; $n_{\mathrm{th}}$
is the mean thermal phonon number, given by the Bose-Einstein
statistics $n_{\mathrm{th}}=[\exp (\hbar \omega _{m}/k_{B}T)-1]^{-1}$ with
the Boltzmann constant $k_{B}$ and the temperature $T$ of the reservoir at
the thermal equilibrium.

\section{Nonreciprocal photon blockade}

\begin{figure}[tbp]
\includegraphics[bb=88 423 455 707, width=8.5 cm, clip]{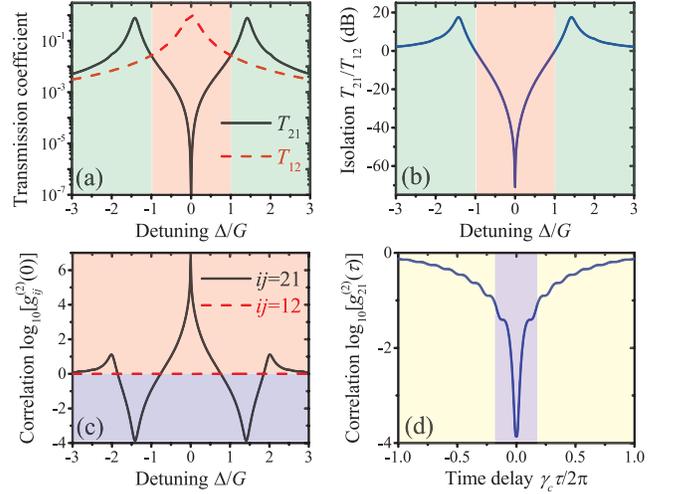}
\caption{(Color online) (a) The transmission coefficients $T_{21}$ (solid
black curve) and $T_{12}$ (dashed red curve) as a function of the detuning $%
\Delta/G$. (b) The isolation as a function of the detuning $\Delta/G$. (c)
The equal-time second-order correlation function $\log_{10}[
g^{(2)}_{ij}(0)] $ ($ij=12,21$) as a function of the detuning $\Delta/G$.
(d) The second-order correlation function $\log_{10}[ g^{(2)}_{21}(\protect%
\tau)]$ as a function of the normalized time delay $\protect\gamma_c \protect%
\tau /2\protect\pi$ at detuning $\Delta=\protect\sqrt{2}G$. The other
parameters are $\Delta _{m}=\Delta/2$, $G=3\protect\gamma_c$, $\protect%
\varepsilon=\protect\gamma_c/10$, $\protect\gamma_m=\protect\gamma_c/100$,
and $n_{\mathrm{th}}=0$.}
\label{fig2}
\end{figure}

In Fig.~\ref{fig2}(a), we show the transmission coefficients $T_{21}$ for
probe field transport from port $1$ to port $2$, and $T_{12}$ for probe
field transport from port $2$ to port $1$. For $T_{21}$, there are two peaks
at $\Delta=\pm \sqrt{2}G$ and one dip at $\Delta=0$. In contrast, there is
only one peak at $\Delta=0$ for $T_{12}$. Figure~\ref{fig2}(b) shows the
isolation $I=T_{21}/T_{21}$ for probe field transport from port $1$ to port $%
2$ as a function of the detuning $\Delta/G$. Isolation for the direction $%
1\rightarrow 2$ is more than $17$ dB at detuning $\Delta=\pm \sqrt{2}G$, and
isolation for the reverse direction of $2\rightarrow 1$ is more than $70$ dB
at detuning $\Delta=0$.

To explore the statistic properties of the transmitted photons, the
equal-time second-order correlation function $\log_{10}[ g^{(2)}_{ij}(0)]$ ($%
ij=12,21$) is shown as a function of the detuning $\Delta/G$ in Fig.~\ref%
{fig2}(c). The photon transport from port $2$ to port $1$ are coherent in
full frequency, i.e., $g^{(2)}_{ij}(0)=1$. The photons transport from port $%
1 $ to port $2$ exhibit strong antibunching effect, i.e., $%
g^{(2)}_{ij}(0)\ll 1 $, around the detunings $\Delta=\pm\sqrt{2}G$, and
exhibit strong bunching effect, i.e., $g^{(2)}_{ij}(0)\gg 1$, around the
detunings $\Delta=0$ and $\pm 2G$. The time duration for nonreciprocal
photon blockade at $\Delta=\pm\sqrt{2}G$ is on the order of $%
2\pi/(10\gamma_{c})$, as shown in Fig.~\ref{fig2}(d).

The peak for $T_{12}\approx 1$ at detuning $\Delta=0$ can be understood by
the fact that when the probe field is injected from port $2$, only the
(linear) optical mode $a_{L}$ can be excited. Thus the maximum transmission
coefficient is reached for the probe field in resonance with the optical
mode $a_{L}$, i.e., $\Delta=0$, and the transmitted photons keep the
statistic properties of the probe field (a coherent field), i.e., $%
g^{(2)}_{12}(0)=1$, for there is no nonlinear interactions in the optical
path from port $2$ to port $1$.

\begin{figure}[tbp]
\includegraphics[bb=27 171 566 632, width=8.5 cm, clip]{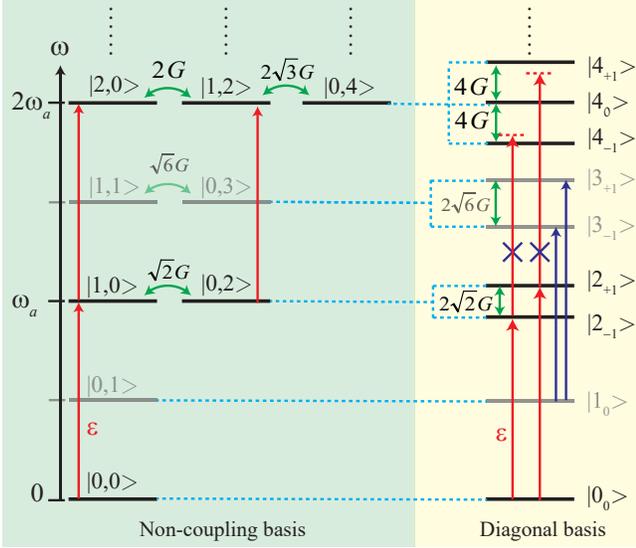}
\caption{(Color online) The schematic energy spectrum of the linearized
quadratically optomechanical coupling between optical mode $a_{L}$ and
mechanical resonator $b$, given in the non-coupling basis (left) and in the
diagonal basis (right).}
\label{fig3}
\end{figure}

In order to understand the origin of the peak for $T_{21}\approx 0.8$ around
the detuning $\Delta =\pm \sqrt{2}G$ and the dip for $T_{21}\approx 10^{-7}$
at $\Delta =0$, we use the ansatz that when the probe field input from port $%
1$, the optical mode $a_{L}$ and the mechanical mode $b$ will be excited, so
the wave function can be written as $|\psi \rangle =C_{00}\left\vert
0,0\right\rangle +C_{10}\left\vert 1,0\right\rangle +C_{02}\left\vert
0,2\right\rangle +\cdots $, as shown in Fig.~\ref{fig3} (left). Here, $%
\left\vert n,m\right\rangle $ represents the Fock state with $n$ photons in $%
a_{L}$ and $m$ phonons in $b$. The wave function can also be written in the
diagonal basis as $|\psi \rangle =C_{0}\left\vert 0_{0}\right\rangle
+C_{1+}\left\vert 2_{+1}\right\rangle +C_{1-}\left\vert 2_{-1}\right\rangle
+\cdots $, as shown in Fig.~\ref{fig3} (right). Under weak probe condition,
the maximum transmission coefficient $T_{21}\approx 0.8 $ is reached for the
probe field in resonance with the transition $\left\vert 0_{0}\right\rangle
\rightarrow \left\vert 2_{\pm 1}\right\rangle $, i.e., $\Delta =\pm \sqrt{2}%
G $. However, the photons absorbed in the transition $\left\vert
0_{0}\right\rangle \rightarrow \left\vert 2_{\pm 1}\right\rangle $ blocks
the transition $\left\vert 2_{\pm 1}\right\rangle \rightarrow \left\vert
4_{\pm 1}\right\rangle $ for large detuning, so we have $g_{12}^{(2)}(0)\ll
1 $ around the detuning $\Delta =\pm \sqrt{2}G$. The dip for $T_{21}\approx
10^{-7}$ at $\Delta =0$ arises from the quantum interference between the
transitions $\left\vert 2_{+ 1}\right\rangle \rightarrow \left\vert
0_{0}\right\rangle $ and $\left\vert 2_{- 1}\right\rangle \rightarrow
\left\vert 0_{0}\right\rangle $, in an equivalent picture as
optomechanically induced transparency~\cite%
{AgarwalPRA10,WeisSci10,SafaviNaeiniNat11}, or electromagnetically induced
transparency in lambda-type three-level atoms~\cite%
{HarrisPT97,FleischhauerRMP05}. Moreover, when $\Delta =0$, the transition $%
\left\vert 0_{0}\right\rangle \rightarrow \left\vert 2_{\pm 1}\right\rangle $
is suppressed, but the two-photon transition $\left\vert 0_{0}\right\rangle
\rightarrow \left\vert 4_{0}\right\rangle $ is resonant, which induces
two-photon tunneling form port $1$ to point $2$, i.e., $g^{(2)}_{21}(0)\gg 1$%
. Similarly, $g^{(2)}_{21}(0)\gg 1$ around $\Delta =\pm 2 G$ is induced by
the resonant transition $\left\vert 0_{0}\right\rangle \rightarrow
\left\vert 4_{\pm}\right\rangle $.

\begin{figure}[tbp]
\includegraphics[bb=106 301 480 598, width=8.5 cm, clip]{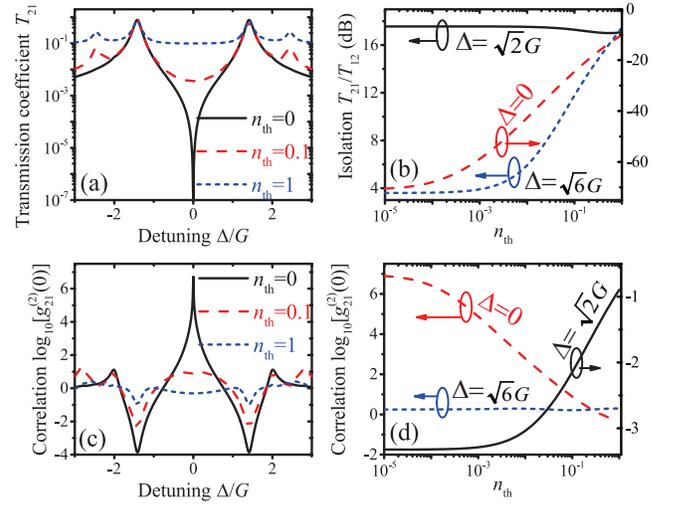}
\caption{(Color online) (a) The transmission coefficient $T_{21}$ and (c)
the equal-time second-order correlation function $\log_{10}[
g^{(2)}_{21}(0)] $ versus the detuning $\Delta/G$ with different mean
thermal phonon number ($n_{\mathrm{th}}=0,0.1,1$). (b) The isolation $%
T_{21}/T_{12}$ and (d) the equal-time second-order correlation function $%
\log_{10}[ g^{(2)}_{21}(0)]$ versus the mean thermal phonon number $n_{%
\mathrm{th}}$ with different detuning ($\Delta=0,\protect\sqrt{2}G,\protect%
\sqrt{6}G$). The other parameters are the same as in Fig.~\protect\ref{fig2}.
}
\label{fig4}
\end{figure}

Before the end of this section, we discuss the effect of thermal phonons on
the nonreciprocal photon blockade. Figures~\ref{fig4}(a) and \ref{fig4}(c)
show the transmission coefficient $T_{21}$ and the equal-time second-order
correlation function $\log_{10}[ g^{(2)}_{21}(0)]$ versus the detuning $%
\Delta/G$ with different mean thermal phonon number ($n_{\mathrm{th}%
}=0,0.1,1 $). Thermal phonons have little influence on the transmission
coefficient $T_{21}$ around $\Delta=\pm\sqrt{2}G$, but have great effect on
the transmission coefficient $T_{21}$ around $\Delta=0$ and $\pm\sqrt{6}G$.
Thermal phonons have little influence on the second-order correlation
function $\log_{10}[ g^{(2)}_{21}(0)]$ around $\Delta=\pm\sqrt{6}G$, but
have great effect on the second-order correlation function $\log_{10}[
g^{(2)}_{21}(0)]$ around $\Delta=0$ and $\pm\sqrt{2}G$.

The relation of the isolation $T_{21}/T_{12}$ and the second-order
correlation function $\log_{10}[ g^{(2)}_{21}(0)]$ on the mean thermal
phonon number $n_{\mathrm{th}}$ are shown in Fig.~\ref{fig4}(b) and \ref%
{fig4}(d) with detuning $\Delta=0,\sqrt{2}G,\sqrt{6}G$. The isolation $%
T_{21}/T_{12}$ around $\Delta=\sqrt{2}G$ is robust against the thermal
phonons, but the antibunching effect of the transport photons become much
weaker for greater thermal phonons. Both the isolation $T_{21}/T_{12}$ and
second-order correlation function $\log_{10}[ g^{(2)}_{21}(0)]$ around $%
\Delta=0$ are sensitive to the mean thermal phonon number $n_{\mathrm{th}}$,
and this quality may be used in accurate temperature measurement at
ultra-low temperature. More interestingly, a peak appears around $\Delta=\pm%
\sqrt{6}G$ in the transmission coefficient $T_{21}$, and the isolation $%
T_{21}/T_{12}$ can be improved with a larger thermal phonon number $n_{%
\mathrm{th}}$. This abnormal effect is induced by the phonon states, e.g., $%
\left\vert 1_{0}\right\rangle$ in Fig.~\ref{fig3}(right). As the temperature
increases, the population probability in $\left\vert 1_{0}\right\rangle$
increases, and the transitions of $\left\vert 1_{0}\right\rangle \rightarrow
\left\vert 3_{\pm 1}\right\rangle $ with resonance frequency $\Delta=\pm%
\sqrt{6}G$ become remarkable gradually, which induces the increasing peaks
of the transmission coefficient $T_{21}$ (or the isolation $T_{21}/T_{12}$)
around $\Delta=\pm\sqrt{6}G$.

Finally, let us discuss the experimental feasibility of our proposal. In
order to the photon correlation induced by the weak probe field, we should
spectrally filter out the strong optical driving field at $\Delta=\Delta_a$
under the condition $\Delta_a \gg \{\gamma_c,G \}$, which has already been
realized in a recent experiment~\cite{CohenNat15}. Another important
condition required to observe nonreciprocal photon blockade in quadratical
coupled optomechanical systems is the well-resolved sideband limit, i.e., $%
\omega_{m}\gg \gamma _{c}$. This requirement may be reached for a high
frequency graphene sheet suspended on WGM microcavities~\cite{HKLiPRA12} or
photonic-crystal microcavities~\cite{WangArx18}.

\section{Conclusions}

In summary, nonreciprocal photon blockade can be realized by directional
nonlinear interactions. We have demonstrated this principle in an
optomechanical system with quadratic optomechanical coupling. We explicitly
show how quadratic optomechanical couplings between two WGM modes and one
mechanical mode are generated when the linear optomechanical couplings
vanishes in the normal optical modes. A quadratic optomechanical system with
WGMs has been used to demonstrate nonreciprocal photon blockade. By
optically pumping the WGM in one direction, the effective quadratic
optomechanical coupling is only enhanced in that direction, and
consequently, the system exhibits nonreciprocal photon blockade. Moreover,
the thermal phonons have important influence on the nonreciprocal photon
blockade, especially on the statistic properties of the transport photons.
Our proposal can have an application to unidirectional single-photon
sources, unidirectional single-photon routers, single-photon isolators and
circulators. This work can also be extended to study phonon manipulation in
double-cavity optomechanics, e.g., nonreciprocal phonon blockade,
nonreciprocal phonon laser~\cite%
{GrudininPRL10,HWangPRA14,HLvPRApp17,YLZhangNJP18,JZhangNPo18},
nonreciprocal photon-phonon entanglement and quantum transfer, etc.

\vskip 2pc \leftline{\bf Acknowledgement}

We thank professor Yu-xi Liu for helpful discussions. X.W.X. is supported by
the National Natural Science Foundation of China (NSFC) under Grants
No.11604096 and the Startup Foundation for Doctors of East China Jiaotong
University under Grant No. 26541059. Y.J.Z. is supported by the China
Postdoctoral Science Foundation under grant No. 2017M620945. H.J. is
supported by NSFC under Grant Nos. 11474087 and 11774086. A.X.C. is
supported by NSFC under Grant No. 11775190.


\bibliographystyle{apsrev}
\bibliography{ref}

\end{document}